\documentclass{article}

\usepackage{PRIMEarxiv}

\usepackage[utf8]{inputenc} 
\usepackage[T1]{fontenc}    
\usepackage{url}            
\usepackage{booktabs}       
\usepackage{amsfonts}       
\usepackage{amsthm}
\usepackage{physics}

\usepackage{nicefrac}       
\usepackage{microtype}      
\usepackage{lipsum}
\usepackage{fancyhdr}       
\usepackage{graphicx}       
\graphicspath{{media/}}     
\usepackage{amsmath,amssymb,amsfonts}
\usepackage{graphicx,psfrag}
\usepackage{enumerate}
\usepackage{url} 
\usepackage{bm}
\usepackage{siunitx}
\usepackage{rotating}
\usepackage{xparse}
\newtheorem{proposition}{Proposition}

\title{Jointly modeling time-to-event and longitudinal data with individual-specific change points: a case study in modeling tumor burden}

\author{
  Ethan M. Alt$^\dagger$, Yixiang Qu$^\dagger$, Emily Damone \\
  {Department of Biostatistics}, \\{University of North Carolina at Chapel Hill}, \\Chapel Hill, NC, USA
   \And
  Jing-ou Liu, Chenguang Wang \\
  Regeneron \\
  Tarrytown, NY, USA\\
  \AND
  Joseph G. Ibrahim \\
  {Department of Biostatistics}, \\{University of North Carolina at Chapel Hill}, \\Chapel Hill, NC, USA\\
  \texttt{ibrahim@bios.unc.edu} \\
}

\begin{document}
\maketitle

\begin{abstract}
In oncology clinical trials, tumor burden (TB) stands as a crucial longitudinal biomarker, reflecting the toll a tumor takes on a patient's prognosis. With certain treatments, the disease's natural progression shows the tumor burden initially receding before rising once more.
Biologically, the point of change may be different between individuals and must have occurred between the baseline measurement and progression time of the patient, implying a random effects model obeying a bound constraint. However, in practice, patients may drop out of the study due to progression or death, presenting a non-ignorable missing data problem. In this paper, we introduce a novel joint model that combines time-to-event data and longitudinal data, where the latter is parameterized by a random change point augmented by random pre-slope and post-slope dynamics. Importantly, the model is equipped to incorporate covariates across for the longitudinal and survival models, adding significant flexibility. Adopting a Bayesian approach, we propose an efficient Hamiltonian Monte Carlo algorithm for parameter inference. We demonstrate the superiority of our approach compared to a longitudinal-only model via simulations and apply our method to a data set in oncology.
\end{abstract}


\section{Introduction}
\label{
sec:intro}
In oncology clinical trials, the gold standard is to measure overall survival (OS) as a primary outcome. However, trials using OS can take a very long time, as statistical power is determined by the number of events (i.e., deaths) rather than the number of enrolled subjects. In order to reduce the burden of recruiting such a large number of subjects, which can often be prohibitive for logistical reasons (e.g., cost), surrogate outcomes have been proposed. The most widely used of these is progression free survival (PFS), which is defined to be the time from enrollment until detection of progression of the disease or death. Thus, PFS is a measurement of tumor burden (TB), albeit a somewhat crude one.

Due to advances in imaging, TB can be measured more precisely and longitudinally. By using a continuous, longitudinal measurement, oncology trials may require less of a sample size than what would be required under a time-to-event (TTE) analysis. Unfortunately, use of TB as an outcome is not straightforward since (1) it is believed that treatments will initially decrease TB before it increases or plateaus over time, (2) in general, the time during which TB begins to increase is bounded from above by the progression time, (3) the change points are specific to the individual, and (4) patients can drop out of the study due to progression, death, or other causes. 

The complex biology of the disease presents several challenging statistical problems. First, there is the issue of modeling patient-specific change points with bound constraints, which presents both serious methodological and computational challenges, as discussed below. Second, if a participant drops out of the study, their event time is unobserved and, hence, we do not observe a bound on the change point. Thus, to model TB data, we need a joint model for longitudinal and survival in which the main focus is to make inferences on the parameters in the longitudinal model with the survival model acting as a non-ignorable missing data mechanism. Joint models for longitudinal and survival data typically focus on inference in the survival portion of the model, in which the longitudinal model is added based on a biomarker which would yield higher precision for the treatment effect in the survival model. 

The literature on these types of joint models is extensive, and a review paper on such models is given in \cite{gould_joint_2015}. In these types of joint models, we model the joint distribution of the longitudinal outcome $y$ and survival time $t$ as $f(y,t) = f(t|y) f(y)$. In our setting, we do an opposite type of conditioning, that is, since the survival model is acting as a missing data mechanism, our joint model for TB, after marginalizing over random effects, is written as $f(y,t) = f(y|t) f(t)$.  

The literature on modeling $f(y,t) = f(y|t) f(t)$ is comparatively sparse, with one well known paper by \cite{hogan_mixture_1997} in which they discretize the survival time and present an application related to dropout in studies of mental health. The statistical literature on modeling TB treating survival time as a non-ignorable missing data mechanism is also quite sparse. A relatively recent paper is \cite{shen_joint_2014} in which they consider such a joint model. However, the longitudinal component of their model is just a standard linear mixed model and does not employ a changepoint. The survival component of their model is a proportional hazards model with a common random effect connecting the longitudinal and survival model.

Regardless of conditioning, most methods that jointly model longitudinal and time-to-event data rely on linear models for the longitudinal part \cite{tsiatis2004joint,rizopoulos2009fully,asar2015joint,brilleman2019joint}. On the other hand, many biological processes do not adhere well to linear models. For example \cite{brilleman_bayesian_2017}, working in a longitudinal-only setting, consider individual-specific change points to model BMI rebound in children. \cite{desmee2017using} use the Stochastic-Approximation Expectation Maximization (SAEM) algorithm to jointly model time-to-event and longitudinal data with nonlinear mixed effects in a prostate cancer setting.
Previous joint time-to-event and longitudinal models have been proposed \cite{altzerinakou2021change,tapsoba2011joint}.
However, none of these aforementioned approaches do not directly address the underlying biology of our application, namely, that the change point must have occurred prior to the progression time.
\cite{kerioui2022modelling} provides an excellent review of joint longitudinal and time-to-event models with nonlinear longitudinal effects.

As far as change point models themselves are concerned, many such models have been proposed in the literature. \cite{van_den_hout_change_2013}, \cite{dominicus_random_2008}, and \cite{hall_bayesian_2003} consider individual change point models for cognitive decline, where elderly patients can experience a drastic shift in cognitive ability at different ages. \cite{lange_hierarchical_1992} and \cite{kiuchi_change_1995} propose an individual-specific change point model to model the progression from HIV to AIDS, where it is believed that T4 cell counts initially decline gradually upon HIV infection, which precedes a sharper decline. However, none of these approaches address dropout and/or censoring, a common component in clinical trials. 

In this paper, we propose a novel individualized change point approach to model TB. Specifically, we propose a random effects model, where each individual has its own change point, intercept, pre-slope (i.e., slope before the change point), and post-slope (i.e., slope after the change point). Because disease progression implies that TB has increased, a bound constraint is imposed on the change point, namely, that the change point must be non-negative and cannot have occurred after the PFS time. Thus, we develop the partially truncated multivariate normal (PTMVN) distribution, which allows one component of a multivariate normal distribution to have bound constraints. Thus, the proposed model adds a much higher layer of complexity and innovation compared to a standard mixed model for the longitudinal component since modeling and estimating the change point is a crucial scientific element for the researcher as it sheds light on the time course of the disease, efficacy of treatment, longevity of the treatment effect, and potential directions for new treatments. Such information is not captured as accurately with standard linear mixed models. 

The rest of this paper proceeds as follows. 
In Section \ref{sec:motivation}, we motivate our model with a Phase 3 oncology trial. 
Section \ref{sec:methods} presents the proposed methodology. 
We provide analytical results for the marginal distribution of the longitudinal outcome in Section \ref{sec:marginal}.
Section \ref{sec:bayes_analysis} discusses Bayesian analysis of the proposed model. 
Section \ref{sec:sims} demonstrates the superiority of our approach compared to ignoring the time-to-event data via a rigorous simulation study. 
In Section \ref{sec:analysis}, we apply our proposed approach to a data set in oncology that mimics a real clinical trial. We briefly conclude in Section \ref{sec:discussion}.

\section{The EMPOWER Study}
\label{sec:motivation}

To motivate our methodology, we consider a recently published phase 3 clinical trial in non-small cell lunch cancer whose initial results were reported in Sezer et al. (2021). The EMPOWER-Lung 1 study (henceforth referred to simply as the EMPOWER study) is a multicentre, open-label, global, phase 3 study exploring the effectiveness of cemiplimab monotherapy for first-line treatment of advanced non-small-cell lung cancer (NSCLC) among smokers aged $\ge 18$ years. 184 enrolled patients were randomly assigned 1:1 to cemiplimab 350 mg every 3 weeks or platinum-doublet chemotherapy, which is the standard of care (SOC). The primary endpoints were OS and PFS, but TB, defined to be the percent change from baseline of tumor size, was also longitudinally recorded. As in most oncology studies, the EMPOWER, used the Response Evaluation Criteria in Solid Tumors (RECIST 1.1) criteria for response evaluation to assess disease. Specifically, disease progression is defined as: i) at least a 20\% increase in the sum of the diameters of target lesions, taking as reference the smallest sum on study (this includes the baseline sum if that is the smallest on study), ii)  an absolute increase of at least 5 mm in the sum of the diameters of target lesions,  iii) the appearance of one or more new lesions, and iv) unequivocal progression of existing non-target lesions.  We refer the reader to Sezer et al. (2021) for further details regarding this trial. 


For the EMPOWER study, there is a priori knowledge that both cemiplimab and the SOC, on average, initially reduce TB before it either plateaus or begins to increase. Such a biological process may be modeled via individual-specific change points. These models allow individuals to have their own intercept, change point, pre-slope, and post-slope. Due to the a priori knowledge about the direction and magnitude of these effects, we propose a Bayesian analysis of the data, where such knowledge may be incorporated via the prior distribution. A Bayesian analysis of the model also avoids the necessity of computing marginal likelihoods (i.e., integrating out the random effects), which can be computationally prohibitive.

Like many other trials, many subjects in the EMPOWER study dropped out or were administratively censored. Moreover, disease progression occurs when, roughly, the TB exceeds a certain threshold. This implies that the change point must have occurred somewhere after baseline but before the time of disease progression, necessitating a joint time-to-event and longitudinal model, where the individual-specific change point obeys a bound constraint. The joint distribution of the random effects (i.e., the change point, intercept, pre-slope, and post-slope) must be able to handle such a bound. 

\begin{figure}
    \centering
    \includegraphics[width=0.9\textwidth]{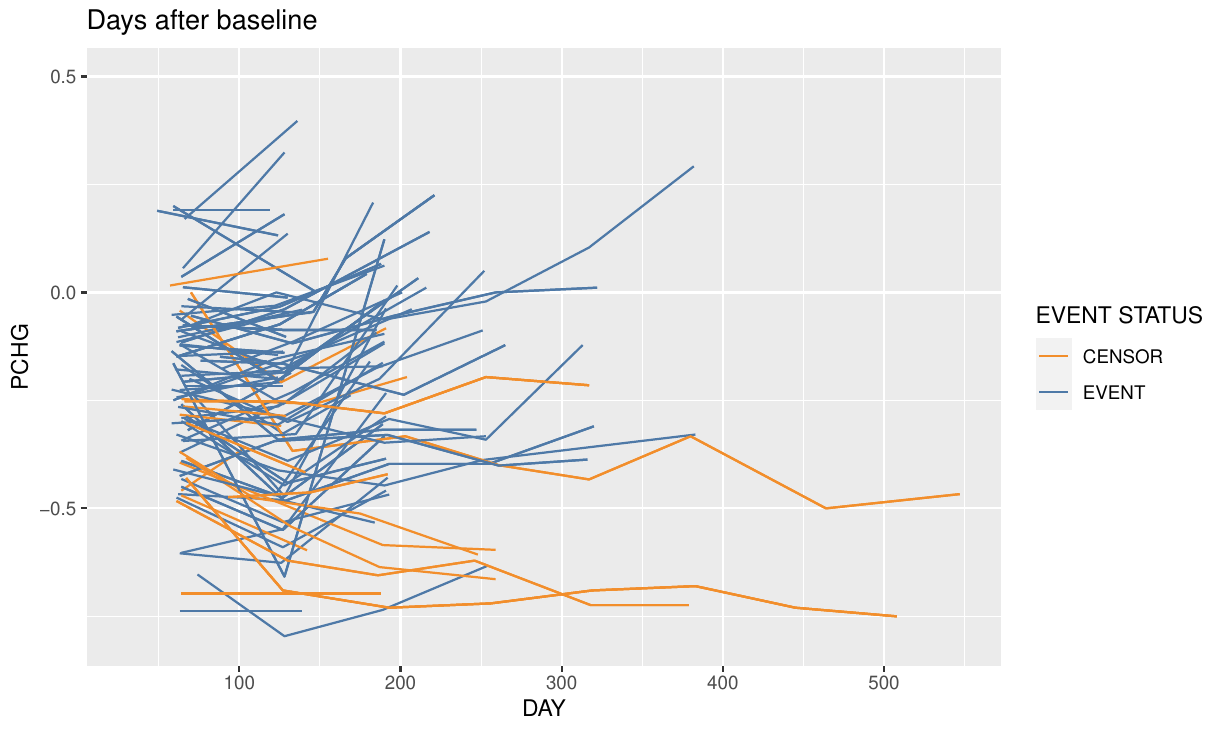}
    \caption{Longitudinal trajectories of tumor burden among the control arm in the simulated EMPOWER study.}
    \label{fig:spaghetti}
\end{figure}

Figure~\ref{fig:spaghetti} depicts the longitudinal trajectories for simulated data designed to mimic the EMPOWER study control arm. Among those who experience the event, it is clear that, in general, tumor burden decreases before it begins to increase. However, the time at which the tumor burden increases varies among patients, suggesting that the change point is individual-specific. The censored subjects present a challenge to the analysis of the data since, for many of them, the longitudinal response tends to be somewhat flat. This may be because the subjects dropped out of the study before their tumor burden increases. Thus, from this figure we see the various complex patterns that emerge from the longitudinal trajectories for each subject and the role of the missing data mechanism for those who experience the event as well as for those who are right censored. It is precisely these phenomena we try to capture in our proposed joint change model that will shed much light on the treatment as well as the time course of the disease. 

Note that Figure~\ref{fig:spaghetti} also shows some individuals whose tumor burden increases from baseline. It is tempting to believe that these individuals may not have been at risk for a change point, implying a change point model would be misspecified for such individuals. However, it is important remember that we do not observe tumor burden continuously. Rather, we only observe its trajectory at finitely many snapshots. Thus, for the individuals whose tumor burden only increases, it is likely to be the case that their change point occurred between the baseline and first clinic visits.

\begin{figure}
    \centering
    \includegraphics[width=0.9\textwidth]{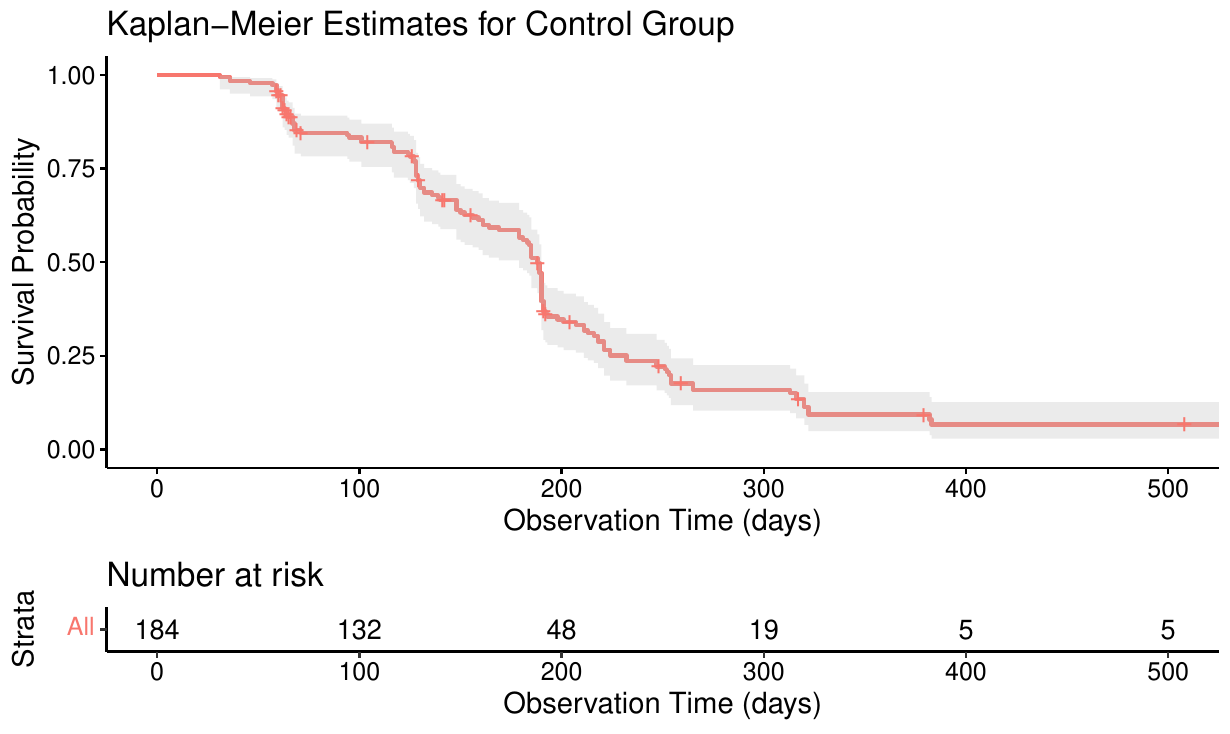}
    \caption{Kaplan-Meier curve for the control arm in the simulated EMPOWER study.}
    \label{fig:km}
\end{figure}

Figure~\ref{fig:km} depicts the Kaplan-Meier 
curve for the control arm of the simulated EMPOWER study. Within the first 250 days after baseline, the survival curve sharply declines. After 250 days, however, the survival curve flattens. The number of subjects beyond this threshold is quite small, but not insignificant. These subjects are likely not at risk for the change point within the window of observation, which presents a statistical modeling challenge that is exacerbated when the subjects are censored. It is these types of modeling challenges that we attempt to address here along with the computational difficulties that they pose. 

It is important to mention here that patients may get disease progression due to  having new lesions.
However, it would be over-complicated to consider disease progression as a mixture of binary (new lesions) and continuous (percent change in the diameters) variables. Thus, in typical oncology clincal practice, when there is disease progression due to new lesions, a pseudo tumor burden measurement is inserted as a 20\% increase from the smallest sum on study. 

Moreover, We note that there is no issue in assuming there is always a positive slope before disease progression, as things must have worsened to be considered as disease progression. On the other hand, we need to make the assumption that there is always a negative slope right after the treatment. Note that patients may not have any observed tumor shrinkage on the study. In such cases, we may assume the change point is close to time 0, but unobserved. As long as there is both a negative and positive slope before disease progression, the change point model is appropriate and  not misspecified.


\section{Methods}
\label{sec:methods}
\subsection{Notation and setup}
Let $t_i = \min\{ t_i^*, c_i^* \}$ denote the observed survival time, where $t_i^*$ is the individual's event time and $c_i^*$ is the individual's censoring time, and let $\nu_i = 1\{ t_i^* \le c_i^* \}$ denote whether individual $i$ experienced the event or was censored. Let $\bm{y}_i = (y_{i1}, \ldots, y_{in_i})'$ denote the longitudinal outcome for individual $i$, where $n_i \ge 1$ is the number of longitudinal measurements recorded for subject $i$. Finally, let $\bm{s}_i = (s_{i1}, \ldots, s_{in_i})'$ denote the times at which the longitudinal measurements were collected. Note that $s_{in_i} \le t_i$ since we do not observe longitudinal outcomes for individuals when they are lost to follow up or experience disease progression or death. 

We assume that the time-to-event and longitudinal outcomes are independent conditional on the random effects. Let the parameters for the models of the time-to-event outcome, random effects, and longitudinal outcome be denoted respectively by $\bm{\theta}_T$, $\bm{\theta}_R$, and $\bm{\theta}_Y$, and let $\bm{\theta} = (\bm{\theta}_T', \bm{\theta}_R', \bm{\theta}_Y')'$. 
Let $f_T(\cdot | \bm{\theta}_T)$, $f_{R|T}(\cdot | t; \bm{\theta}_R)$, and $f_{Y|R}(\cdot | \bm{r}; \bm{\theta}_Y)$ denote arbitrary parametric densities for the time-to-event outcome, random effects (conditional on the event time), and the longitudinal outcomes (conditional on the random effects), respectively. Finally, let $S_T(t | \bm{\theta}_T) = \int_t^{\infty} f_T(s | \bm{\theta}_T) ds$ denote the survival function corresponding to $f_T(\cdot | \bm{\theta}_T)$.

In this paper, we propose a piecewise linear mixed model for the longitudinal outcomes with a random change point, i.e.,
\begin{align}
    y_{ij} = \bm{x}_{ij}'\bm{\beta} + b_{0i} + b_{1i}(s_{ij} - \omega_{i})1\{ s_{ij} \le \omega_i \} + b_{2i}(s_{ij} - \omega_i)1\{ s_{ij} > \omega_i \} + \epsilon_{ij}, \ \ j = 1, \ldots, n_i
\end{align}
where $\bm{x}_{ij}$ is a matrix of covariates for subject $i$ at visit $j$ pertaining to fixed effects (which may include an intercept term), $\bm{\beta}$ is a vector of regression coefficients, $b_{0i}$ is a random intercept, $b_{1i}$ is a random slope prior to the change point $\omega_i$, $b_{2i}$ is a random slope after the change point, and $\bm{\epsilon}_i \sim N_{n_i}(0, \bm{\Sigma}_{\bm{y}_i | \bm{r}_i})$ is an error term. We assume $\bm{\Sigma}_{\bm{y}_i | \bm{r}_i} = \sigma^2_y \bm{I}_{n_i}$, where $\sigma^2_y > 0$ and $\bm{I}_n$ is the $n$-dimensional identity matrix. The constraints on $\omega_{i}$ are a necessary part of the model biologically as they play a fundamental role in modeling the change point itself. Specifically, our proposed model puts the boundary of the change point at the ``true'' progression time. Thus, early dropout and administrative censoring are handled by what me be termed ``progression data augmentation.'' Finally, we mention that Thomadakis et al. \cite{thomadakis2019longitudinal} showed that joint models may be biased in situations where the time to event is based on observed value above a threshold and the study dropout mechanism is assumed to be missing at random (MAR). However, our approach does 
{\em not} assume MAR and in fact models the dropout mechanism through a shared frailty survival model, thus deeming the dropout mechanism to be nonignorably missing as discussed in detail in Section 1.

Let $\bm{r}_i = (\bm{b}_i', \omega_i)'$ denote the random effects for subject $i$. If individual $i$ experiences the event (i.e., if $\nu_i = 1$), then $t_i^* = t_i$ and the individual's likelihood contribution is proportional to
\begin{align}
    f(t_i^* = t_i, \bm{y}_i | \bm{\theta}) = \int_{0}^{t_i} \int_{\mathbb{R}^3} f_T(t_i | \bm{\theta}_T) f_{R|T}(\bm{r}_i | t_i; \bm{\theta}_R) f_{Y|R}(\bm{y}_i | \bm{r}_i; \bm{\theta}_Y)
    d\bm{b}_i d\omega_i
    \label{eq:obs_like}
    ,
\end{align}
where, for subject $i$, $\bm{b}_i$ is a vector of random regression coefficients and $\omega_i > 0$ is a change point. Conversely, if individual $i$ was censored (i.e., if $\nu_i = 0$), then we only know $t_i^* > t_i$ and the individual's contribution to the likelihood is proportional to
\begin{align}
    f(t_i^* > t_i, \bm{y}_i | \bm{\theta}) 
    = \int_{t_i}^{\infty} \int_{0}^{t} \int_{\mathbb{R}^3}
    f_T(t | \bm{\theta}_T) f_{R|T}(\bm{r}_i | t, \bm{\theta}_R) f_{Y|R}(\bm{y}_i | \bm{b}_i, \omega_i; \bm{\theta}_Y)
    d\bm{b}_i d\omega_i dt
    %
    .
    \label{eq:cens_like}
\end{align}

Let $n_c$ denote the number of censored patients. Assuming without loss of generality that the first $n_c$ patients are censored, the complete data likelihood (i.e., the likelihood if we observed all event times and random effects) is given by
\begin{align}
    L^c(\bm{\theta} | D^c) &\propto 
        \prod_{i=1}^{n_c} f_T(t_i^* | \bm{\theta}_T) f_{R|T}( \bm{r}_i | t_i^*, \bm{\theta}_R) f_{Y|R}(\bm{y}_i | \bm{r}_i, \bm{\theta}_Y)
        I(t_i^* > t_i)
    \notag \\
    &\hspace{1cm} \times
        \prod_{i=n_c + 1}^{n} f_T(t_i | \bm{\theta}_T) f_{R|T}( \bm{r}_i | t_i, \bm{\theta}_R) f_{Y|R}(\bm{y}_i | \bm{r}_i, \bm{\theta}_Y),
    \label{eq:like_complete}
    .
\end{align} 

Using \eqref{eq:like_complete}, the observed data likelihood is given by
\begin{align}
    L(\bm{\theta} | D) &\propto
        \left[ \prod_{i=1}^{n_c} \int_{t_i}^{\infty} \int_0^{t_i^*} \int_{\mathbb{R}^3} f_T(t_i^* | \bm{\theta}_T) f_{R|T}( \bm{r}_i | t_i^*, \bm{\theta}_R) f_{Y|R}(\bm{y}_i | \bm{r}_i, \bm{\theta}_Y) d\bm{b}_i d\omega_i dt_i^* \right]
        \notag\\
        &\times \left[ \prod_{i=n_c + 1}^{n} \int_0^{t_i} \int_{\mathbb{R}^3} f_T(t_i | \bm{\theta}_T) f_{R|T}( \bm{r}_i | t_i, \bm{\theta}_R) f_{Y|R}(\bm{y}_i | \bm{r}_i, \bm{\theta}_Y) d\bm{b}_i d\omega_i \right]
    .
    \label{like_observed}
\end{align}

Typically, a patient progresses when the patient's tumor burden increases. Thus, we impose the constraint that the change point must have occurred no later than the time of progression, i.e., $0 < \omega_i \le t_i^*$. We assume that the density of the random effects is given by
\begin{align}
    f_{R|T}( \bm{r}_i | t_i^* = t_i) \propto \phi_4\left( \bm{r}_i | \bm{\mu}_r, \bm{\Sigma}_r \right) 1\{ \omega_i \le t_i \}
      %
      \label{eq:randeff_dist_observed_mvn}
\end{align}
where $\phi_n(\cdot | \bm{\mu}, \bm{\Sigma})$ is the $n$-dimensional multivariate normal distribution with mean $\bm{\mu}$ and covariance matrix $\bm{\Sigma}$, 
$\bm{\mu}_r = (\bm{\mu}_b', \mu_{\omega})'$ and 
$\bm{\Sigma}_r = 
    \begin{pmatrix} 
        \bm{\Sigma}_{\bm{b}}       & \bm{\sigma}_{\bm{b}\omega} \\
        \bm{\sigma}_{\bm{b}\omega}' & \sigma_{\omega}^2
    \end{pmatrix}
$ 
are, respectively, the mean and covariance matrix of the random effects prior to truncation. We refer to the distribution corresponding to the kernel in \eqref{eq:randeff_dist_observed_mvn} as the ``partially truncated multivariate normal'' (PTMVN) distribution.


\section{Properties of the PTMVN distribution}
\label{sec:ptmvn}
In this section, we present some properties of the partially truncated multivariate normal (PTMVN) distribution. All proofs are provided in Web Appendix A, where more properties of the PTMVN distribution are also derived.

Throughout this section, we denote $\bm{r} = (\bm{b}', \omega)' \sim \text{PTMVN}_q\left( \bm{\mu}, \bm{\Sigma}, l, u \right)$ to mean the $q$-dimensional vector $\bm{r}$ has the PTMVN distribution where $\omega \in [l, u]$ and (the unbounded version of) the vector has multivariate mean and covariance $\bm{\mu}$ and $\bm{\Sigma}$, respectively. Let $\bm{\mu}_{\bm{b} | \omega}$ and $\bm{\Sigma}_{\bm{b} | \omega}$ respectively denote the conditional mean and covariance of $\bm{b}$ given $\omega$. Also, let $\mu_{\omega | \bm{b}}$ and $\sigma^2_{\bm{\omega} | \bm{b}}$ denote the conditional mean and variance of $\omega$ given $\bm{b}$. 

\begin{proposition}
    \label{prop:ptmvn_density}
        The density of $\bm{r}$ is given by
    \begin{align}
        f(\omega, \bm{b}) = \frac{\phi(\bm{r} | \bm{\mu}, \bm{\Sigma})}{\Phi(\upsilon) - \Phi(\lambda)} 1\{ l < \omega < u \},
        \label{eq:ptmvn_joint}
    \end{align}
    where $\lambda = (l - \mu_{\omega}) / \sigma_{\omega}$ and let $\upsilon = (u - \mu_{\omega}) / \sigma_{\omega}$.
\end{proposition}

\begin{proposition}
    \label{prop:ptmvn_conditionals}
    Let $\mu_{\omega | \bm{b}} = \mu_{\omega} + \bm{\sigma_{\bm{b}\omega}}' \bm{\Sigma}_{\bm{b}}^{-1}(\bm{b} - \bm{\mu}_{\bm{b}})$ and 
    $
        \sigma_{\omega | \bm{b}} = \sigma_{\omega}^2 - \bm{\sigma}_{\bm{b}\omega}' \bm{\Sigma}_{\bm{b}}^{-1} \bm{\sigma}_{\bm{b}\omega}
    $
    respectively denote the conditional mean and variance of $\omega$ on the untruncated scale, and let
    $
        \bm{\mu}_{\bm{b} | \omega} = \bm{\mu}_{\bm{b}} + \frac{\bm{\sigma}_{\bm{b}\omega}}{\sigma_{\omega}^2}(\omega - \mu_{\omega})
    $
    and 
    $
        \bm{\Sigma}_{\bm{b} | \omega} = \bm{\Sigma}_{\bm{b}} - \frac{1}{\sigma_{\omega}^2} \bm{\sigma}_{\bm{b}\omega} \bm{\sigma}_{\bm{b}\omega}'
    $
    respectively denote those for $\bm{b}$.
    Then $\omega | \bm{b} \sim \text{TN}(\mu_{\omega | \bm{b}}, \sigma_{\omega | \bm{b}}^2, l, u)$ and $\bm{b} | \omega \sim N_{q-1}(\bm{\mu}_{\bm{b} | \omega}, \bm{\Sigma}_{\bm{b} | \omega}$).
\end{proposition}
\noindent Although this result appears obvious, it is invaluable for generating random variables from the PTMVN distribution when combined with Propostion \ref{prop:ptmvn_marginals} presented next.

\begin{proposition}
    \label{prop:ptmvn_marginals}
    The marginal distributions of the PTMVN distribution are given by $\omega \sim \text{TN}(\mu_{\omega}, \sigma_{\omega}, l, u)$ and
    \begin{align}
        f(\bm{b}) = \frac{ \Phi(\upsilon_{\bm{b}}) - \Phi(\lambda_{\bm{b}}) }{\Phi(\upsilon) - \Phi(\lambda)} \phi(\bm{b} | \bm{\mu}_{\bm{b}}, \bm{\Sigma}_{\bm{b}}),
        \label{eq:ptmvn_b_marginal}
    \end{align}
    where $\upsilon_{\bm{b}} = (u - \mu_{\omega | \bm{b}}) / \sigma_{\omega | \bm{b}}$ and $\lambda_{\bm{b}} = (l - \mu_{\omega | \bm{b}}) / \sigma_{\omega | \bm{b}}$.
    The density in \eqref{eq:ptmvn_b_marginal} is not a recognizable density, although it does integrate to unity.
\end{proposition} 
\noindent Note that combining Propositions \ref{prop:ptmvn_conditionals} and \ref{prop:ptmvn_marginals} implies a direct sampling scheme for generating random varites from the PTMVN distribution. Specifically, we can write $f(\omega, \bm{b}) = f(\omega) f(\bm{b} | \omega)$, where the first quantity is a truncated normal density and the second quantity is a multivariate normal density. 

\begin{proposition}
\label{prop:ptmvn_mgf}
    Let $\bm{t} = (t_{\omega}, \bm{t}_{\bm{b}})' \in \mathbb{R}^{q}$. The moment generating function (MGF) of the PTMVN distribution is given by
    \begin{align}
        M(\bm{t}) =
        %
        \frac{
        C(\bm{t})
    }
    {
        \Phi(\upsilon) - \Phi(\lambda)
    }
    M_N(\bm{t}),
    \label{eq:ptmvn_mgf}
    \end{align}
    where $
        C(\bm{t}) = \Phi\left( \upsilon - \bm{t}'\bm{q} \right)
        -
        \Phi\left( \lambda - \bm{t}'\bm{q} ) \right)
    $,
    $\bm{q} = \bm{\sigma}_{\circ 1} / \sigma_{\omega}$ and where $\bm{\sigma}_{\circ 1}$ is the first column of $\bm{\Sigma}$,
    and 
    $
        M_N(\bm{t}) = \exp\left\{ \bm{t}'\bm{\mu} + \frac{1}{2} \bm{t}'\bm{\Sigma}\bm{t} \right\}
    $ 
    is the MGF of the multivariate normal distribution.
\end{proposition}
Using this result, we show in Web Appendix A (along with the derivation of the covariance) that the mean of the PTMVN distribution is given by
\begin{align}
    E_{\bm{\theta}}\left[ \bm{r} \right] &= \bm{\mu} - \frac{\phi(\upsilon) - \phi(\lambda)}{\Phi(\upsilon) - \Phi(\lambda)} \bm{q}.
    \label{eq:ptmvn_mean}
\end{align}

\section{Marginal distribution of the longitudinal outcome}
\label{sec:marginal}
In this section, we present results for the marginal distribution for the longitudinal outcome. Detailed derivations are provided in Web Appendix B.

Let $\bm{y}_i = (y_{i1}, \ldots, y_{i,n_i})'$ and $\bm{X}_i = (\bm{x}_{i1}, \ldots, \bm{x}_{i,n_i})'$. Let $\Delta_{ij} = s_{ij} - \omega_i$ and let $\bm{z}_{ij} = (1, \Delta_{ij} 1\{ \Delta_{ij} \le 0 \}, \Delta_{ij} 1\{ \Delta_{ij} \ge 0 \})'$. Moreover, let $\bm{Z}_i = (\bm{z}_{i1}, \ldots, \bm{z}_{i,n_i})'$ denote the ``design matrix'' for the random effects $\bm{b}_i$, which itself depends on a random effect $\omega_i$. In particular, we may write
\begin{align}
    \bm{Z}_i = 
    \begin{pmatrix}
        \bm{z}_{i1}' \\ \vdots \\ \bm{z}_{i,n_i}'
    \end{pmatrix}'
    =
    \begin{pmatrix} 
        1 
            & \Delta_{i1} 1\{ \Delta_{i1} \le 0 \}
            & \Delta_{i1} 1\{ \Delta_{i1} \ge 0 \}
        \\
        \vdots & \vdots & \vdots
        \\
        1 
            & \Delta_{i,n_i} 1\{ \Delta_{i,n_i} \le 0 \}
            & \Delta_{i,n_i} 1\{ \Delta_{i,n_i} \ge 0 \}
    \end{pmatrix}
\end{align}
If the event time is observed, the conditional distribution for the longitudinal outcome is given by
$
    \bm{y}_i | \bm{r}_i \sim N_{n_i}(\bm{X}_i \bm{\beta} + \bm{Z}_i \bm{b}_i, \sigma^2_y \bm{I}_{n_i}).
$

Using Proposition \eqref{prop:ptmvn_conditionals}, it follows that
\begin{align}
    \left. \begin{pmatrix}
        \bm{y}_i \\ \bm{b}_i
    \end{pmatrix}
    \right| \omega_i
    \sim
    N_{n_i + 3}\left( 
        \begin{pmatrix}
            \bm{\mu}_{\bm{y} | \omega_i } \\
            \bm{\mu}_{\bm{b} | \omega_i}
        \end{pmatrix}
    ,
    \begin{pmatrix}
        \bm{\Sigma}_{\bm{y} | \omega_i} & \bm{\Sigma}_{\bm{y}\bm{b} | \omega_i} \\
        \bm{\Sigma}_{\bm{y}\bm{b} | \omega_i}' & \bm{\Sigma}_{\bm{b}|\omega}
    \end{pmatrix}
    \right)
    ,
\end{align}
where 
$\bm{\mu}_{\bm{y} | \omega_i} = \bm{X}_i\bm{\beta} + \bm{Z}_i \bm{\mu}_{\bm{b} | \omega_i}$, 
$\bm{\Sigma}_{\bm{y} | \omega_i} = \sigma^2_y \bm{I}_{n_i} + \bm{Z}_i \bm{\Sigma}_{\bm{b} | \omega} \bm{Z}_i'$, and $\bm{\Sigma}_{\bm{y}\bm{b} | \omega} = \bm{Z}_i \bm{\Sigma}_{\bm{b} | \omega}$. Hence, we have that $\bm{y}_i | \omega_i \sim N_{n_i}\left( \bm{X}_i \bm{\beta} + \bm{Z}_i \bm{\mu}_{\bm{b} | \omega_i}, \sigma^2_y \bm{I}_{n_i} + \bm{Z}_i \bm{\Sigma}_{\bm{b} | \omega} \bm{Z}_i' \right)$. When $t_i = t^*_i$, the marginal density of $\bm{y}_i$ is thus given by
\begin{align}
    f(\bm{y}_i) = \int \phi_{n_i}(\bm{y}_i | \bm{X}_i\bm{\beta} + \bm{Z}_i \bm{\mu}_{\bm{b} | \omega_i}, \sigma^2_y \bm{I}_{n_i} + \bm{Z}_i \bm{\Sigma}_{\bm{b} | \omega_i} \bm{Z}_i')
    f_{\text{TN}}(\omega_i | \mu_{\omega}, \sigma_{\omega}^2, 0, t_i^*)
    d\omega_i
    .
\end{align}

Given $t_i = t_i^*$, the marginal mean of $\bm{y}_i$ may be computed as
\begin{equation}
    E\left[ \bm{y}_i \right]
    = E\left[ E\left[ \bm{y}_i | \omega_i \right] \right]
    = \bm{X}_i \bm{\beta} + E\left[ \bm{Z}_i \bm{\mu}_{\bm{b} | \omega_i} \right], 
    \label{eq:y_marginalmean}
\end{equation}
where
\begin{align}
    E\left[ \bm{Z}_i \bm{\mu}_{\bm{b} | \omega_i} \right] = E\left[ \bm{Z}_i \left( \bm{\mu}_{\bm{b}} + \frac{\omega_i - \mu_{\omega}}{\sigma_{\omega}^2} \bm{\sigma}_{\omega \bm{b}} \right)
    \right]
    =
    E[\bm{Z}_i] \left( \bm{\mu}_{\bm{b}} - \frac{\mu_{\omega}}{\sigma_{\omega}^2} \bm{\sigma}_{\omega \bm{b}} \right)
    + E\left[ \omega_i \bm{Z}_i \right] \frac{\bm{\sigma}_{\omega \bm{b}}}{\sigma_{\omega}^2}.
    \label{eq:expectation_linpred}
\end{align}

Expectations of $\bm{y}_i$ involve expectations of $\bm{Z}_i$, which depend on moments of the truncated normal random variable $\omega$. In Web Appendix A, we prove various properties of the PTMVN distribution, including that
\begin{align}
    &E\left[ \omega_i^m \Delta_{ij}^k 1\{  \alpha \le \Delta_{ij} \le \beta \} \right]
    \notag \\
    &=  \frac{ 
        \Phi\left( \frac{s_{ij} - \alpha - \mu_{\omega}}{\sigma_{\omega}} \right)
        - \Phi\left( \frac{s_{ij} - \beta - \mu_{\omega}}{\sigma_{\omega}} \right)
    }{
          \Phi(\frac{b - \mu_{\omega}}{\sigma_{\omega}})
        - \Phi(\frac{a - \mu_{\omega}}{\sigma_{\omega}})
    }
    \sum_{l=0}^k \binom{k}{l} (-1)^l s_{ij}^{k-l}
    m_{\text{TN}}^{(m+l)}(\mu_{\omega}, \sigma_{\omega}^2,  s_{ij} - \beta, s_{ij} - \alpha),
    \label{eq:omega_delta_expectation}
\end{align}
where $m_{TN}^{(k)}(\mu, \sigma^2, a, b)$ is the $k^{th}$ moment of the truncated normal distribution with mean $\mu$, variance $\sigma^2$, and truncation parameters $-\infty \le a < b \le \infty$, which is available via a recursive formula \cite{orjebin_recursive_2014}. We may substitute \eqref{eq:omega_delta_expectation} in the expression \eqref{eq:expectation_linpred} to compute the marginal mean of $\bm{y}_i$ given in \eqref{eq:y_marginalmean}.

Given $t_i = t_i^*$, the marginal covariance matrix of $\bm{y}_i$ is given by
\begin{align}
    \text{Cov}(\bm{y}_i) = \sigma^2_y \bm{I}_{n_i} + E\left[ \bm{Z}_i \left( \bm{\Sigma}_{\bm{b}} - \bm{\sigma}_{\omega \bm{b}} \bm{\sigma}_{\omega \bm{b}}' / \sigma_{\omega}^2 \right) \bm{Z}_i' \right].
    \label{eq:y_margcov}
\end{align}
It is difficult to compute the expectation in \eqref{eq:y_margcov} analytically, but estimates of the covariance matrix may be easily obtained using Monte Carlo techniques.

\section{Bayesian analysis of the model}
\label{sec:bayes_analysis}
In this section, we discuss Bayesian analysis of the proposed joint model. We present full conditional distributions and discuss MCMC sampling algorithms. We also provide recommendations for prior elicitation.

In the Bayesian paradigm, missing data (e.g., the censored event times and the random effects) are treated as parameters. That is, they are sampled. The joint posterior density of the parameters and missing data is proportional to the complete data likelihood and a prior for $\bm{\theta} = (\bm{\theta}_T, \bm{\theta}_R, \bm{\theta}_Y)$, i.e.,
\begin{align}
    p(\bm{t}^*_{\text{cen}}, \bm{\theta}_T, \bm{\theta}_R, \bm{\theta}_Y, \bm{r} | D)
        \propto L(\bm{\theta}_T | \bm{t}^*_{\text{cen}}, D) L(\bm{\theta}_R | \bm{\omega}, \bm{b}, \bm{t}^*) L(\bm{\theta}_Y | D, \bm{\omega}, \bm{b})\pi(\bm{\theta}_T, \bm{\theta}_R, \bm{\theta}_Y),
\end{align}
where $\pi(\bm{\theta}_T, \bm{\theta}_R, \bm{\theta}_Y)$ is a joint prior for $\bm{\theta}_T$, $\bm{\theta}_R$ and $\bm{\theta}_Y$.

\subsection{Full conditional densities}
The full conditional posterior density for the time-to-event parameters is given by
$$
   p(\bm{\theta}_T | D, \bm{t}^*_{\text{cen}}, \bm{r}, \bm{\theta}_R, \bm{\theta}_Y) \propto
   \left[\prod_{i=1}^n f_T(t_i^* | \bm{\theta}_T)\right] \pi(\bm{\theta}_T | \bm{\theta}_R, \bm{\theta}_Y).
$$
Note that if $t^*_i$ is observed for every $i$ and if $\bm{\theta}_T$ is a priori independent of $\bm{\theta}_R$ and $\bm{\theta}_Y$, then $\bm{\theta}_T$ is independent of $\bm{\theta}_R$, $\bm{\theta}_Y$ a posteriori. However, when at least one event time is censored, the parameters will be correlated in general. 

The full conditional posterior density of $\bm{t}^*_{\text{cen}}$ is given by
\begin{align}
    p(\bm{t}_{\text{cen}}^* | D, \bm{r}, \bm{\theta})
    \propto
    \prod_{i=1}^{n_c}
    \frac{ f_T(t_i^* | \bm{\theta}_T) }{\Phi\left( \frac{t_i^* - \mu_{\omega}}{\sigma_{\omega}} \right) - \Phi\left( -\frac{\mu_{\omega}}{\sigma_{\omega}} \right)}
    1\{ t_i^* > t_i \}
    .
    \label{eq:censtimes_fullcond}
\end{align}
Since the full conditional posterior density depends on both $\bm{\theta}_T$ and $\bm{\theta}_R$, the censoring times induce correlation a posteriori. This correlation ultimately feeds into $\bm{\theta}_Y$ through the conditional distribution of the $\bm{y}$ given the random effects $\bm{r}$. Note that the full conditional distribution for the censored event times is \emph{not} simply a failure time distribution truncated from below at the censoring times since we must have $t_i^* \ge \omega_i$.


The full conditional posterior density of $\bm{\theta}_R$ is given by
\begin{align}
    p(\bm{\theta}_R | D, \bm{t}^*_{\text{cen}}, \bm{r}, \bm{\theta}_T, \bm{\theta}_Y)
    \propto 
        \left[ 
            \prod_{i=1}^{n} 
            \frac{ \phi\left( \bm{r}_i | \bm{\mu}_R, \bm{\Sigma}_R \right) }{ \Phi\left( \frac{t_i^* - \mu_{\omega}}{\sigma{\omega}}
            \right)
            - 
            \Phi\left(- \frac{\mu_{\omega}}{\sigma{\omega}}
            \right)
            }
        \right]
        \pi(\bm{\theta}_R | \bm{\theta}_T, \bm{\theta}_Y)
    \label{eq:post_thetaR_fullcond}
\end{align}
Note that, even if $\bm{\theta}_R$ is independent of $(\bm{\theta}_T, \bm{\theta}_Y)$ a priori, it is nontrivial to sample from the full conditional in \eqref{eq:post_thetaR_fullcond} due to the normalizing constant present in the PTMVN density. Specifically, the quantity $\prod_{i=1}^n \left[ \Phi\left( (t_i^* - \mu_{\omega}) / \sigma_{\omega} \right) - \Phi( -\mu_{\omega} / \sigma_{\omega}) \right]^{-1}$ makes it difficult to find a conjugate sampler for $\bm{\theta}_R$. As a result, we propose to use the efficient Hamiltonian Monte Carlo algorithm implemented in the \verb|stan| programming language \cite{carpenter_stan_2017}.

The full conditional posterior density of the random effects is given by
\begin{align}
    p(\bm{\omega},\bm{b} | D, \bm{\theta}_T, \bm{\theta}_R, \bm{\theta}_Y, \bm{t}^*_{\text{cen}})
    \propto
    \prod_{i=1}^{n} 
            \frac{ \phi\left( \bm{r}_i | \bm{\mu}_R, \bm{\Sigma}_R \right) }{ \Phi\left( \frac{t_i^* - \mu_{\omega}}{\sigma_{\omega}}
            \right)
            - 
            \Phi\left(- \frac{\mu_{\omega}}{\sigma_{\omega}}
            \right)
            }
    \phi\left( \bm{y}_i | \bm{X}_i\bm{\beta}_i + \bm{Z}_i(\omega_i) \bm{b}_i, \sigma^2 \bm{I}_{n_i} \right),
\end{align}
which is difficult to sample from directly since $\omega_i, \bm{b}_i$ are correlated in general. On the other hand, using Proposition \ref{prop:ptmvn_conditionals}, it follows that
\begin{align}
    \left. \begin{pmatrix}
        \bm{y}_i \\ \bm{b}_i
    \end{pmatrix}
    \right| \omega_i
    \sim
    N_{n_i + 3}\left( 
        \begin{pmatrix}
            \bm{\mu}_{\bm{y} | \omega_i } \\
            \bm{\mu}_{\bm{b} | \omega_i}
        \end{pmatrix}
    ,
    \begin{pmatrix}
        \bm{\Sigma}_{\bm{y} | \omega_i} & \bm{\Sigma}_{\bm{y}\bm{b} | \omega_i} \\
        \bm{\Sigma}_{\bm{y}\bm{b} | \omega_i}' & \bm{\Sigma}_{\bm{b}|\omega}
    \end{pmatrix}
    \right)
    ,
\end{align}
$\bm{\mu}_{\bm{y} | \omega_i} = \bm{X}_i\bm{\beta} + \bm{Z}_i \bm{\mu}_{\bm{b} | \omega_i}$, 
$\bm{\Sigma}_{\bm{y} | \omega_i} = \sigma^2_y \bm{I}_{n_i} + \bm{Z}_i \bm{\Sigma}_{\bm{b} | \omega} \bm{Z}_i'$, and $\bm{\Sigma}_{\bm{y}\bm{b} | \omega} = \bm{Z}_i \bm{\Sigma}_{\bm{b} | \omega}$. Using properties of the multivariate normal distribution, we have
$
  \bm{b}_i | \bm{y}_i, \omega_i \sim N_3\left(
        \mu_{\bm{b} | \omega_i, \bm{y}_i}, \bm{\Sigma}_{\bm{b} | \omega_i, \bm{y}_i}
  \right),
$
where
\begin{align*}
    \bm{\mu}_{\bm{b} | \omega_i, \bm{y}_i} &= \bm{\mu}_{\bm{b} | \omega_i} + \bm{\Sigma}_{\bm{by} | \omega_i} \bm{\Sigma}_{\bm{y} | \omega_i}^{-1} \left( \bm{y}_i - \bm{\mu}_{\bm{y} | \omega_i} \right), \notag \\
    \bm{\Sigma}_{\bm{b} | \omega_i, \bm{y}_i}
    &= \bm{\Sigma}_{\bm{b} | \omega} - \bm{\Sigma}_{\bm{b} \bm{y}_i | \omega_i} \bm{\Sigma}_{\bm{y}_i | \omega_i} \bm{\Sigma}_{\bm{b}\bm{y}_i | \omega_i}'.
\end{align*}
Thus, one can sample from the full conditional of $\bm{b}_i$ via direct Monte Carlo using a multivariate normal density. However, such sampling requires the calculation of the inverse of $n_i$ matrices each of dimension $3 \times 3$, which can become burdensome computationally. By proposition \eqref{prop:ptmvn_conditionals}, the full conditional density of $\omega_i$ is given by
\begin{align}
    p(\bm{\omega} | D, \bm{\theta}, \bm{b}, \bm{t}_{\text{cen}}^*) \propto
    \prod_{i=1}^n
    \frac{\phi(\omega_i | \mu_{\omega | \bm{b}_i}, \sigma^2_{\omega | \bm{b}}, 0, t_i^*)}{ \Phi\left( \frac{t_i^* - \mu_{\omega} | \bm{b}_i}{\sigma_{\omega | \bm{b}_i}} \right) - \Phi\left( -\frac{\mu_{\omega} | \bm{b}_i}{\sigma_{\omega | \bm{b}_i}} \right) }
    \phi\left( \bm{y}_i | \bm{X}_i \bm{\beta} + \bm{Z}_i(\omega_i) \bm{b}_i, \sigma^2 \bm{I}_{n_i} \right),
\end{align}
which cannot be expressed in terms of a recognizable density. 

The full conditional posterior density of the outcome parameters is given by
\begin{align}
    p(\bm{\theta}_Y | D, \bm{\theta}_T, \bm{\theta}_R, \bm{r}, \bm{t}_{\text{cen}}^*)
    \propto
    \left[ \prod_{i=1}^n \phi(\bm{y}_i | \bm{X}_i \bm{\beta} + \bm{Z}_i \bm{b}_i, \sigma^2 \bm{I}_{n_i}) \right] \pi(\bm{\theta}_Y).
\end{align}
It is easy to show that a normal-inverse-gamma prior on $\bm{\theta}_Y = (\bm{\beta}, \bm{\sigma}^2)$, results in conjugate full conditional posterior density. Nevertheless, due to the computational complexity of the random effects, we propose a more efficient algorithm described below.

\subsection{Efficient sampling}
The random effects can be difficult to compute due to the truncation of the first component and due to their correlation. In this section, we describe how the random effects may be sampled efficiently in a HMC scheme.

By Propositions \ref{prop:ptmvn_conditionals} and \ref{prop:ptmvn_marginals}, may write the joint density of $(\omega_i, \bm{b}_i)$ as the product of a truncated normal density and a conditional multivariate normal density. Let $\bm{L}$ denote the lower Cholesky factor of $\bm{\Sigma}_R$. If $\omega_i$ were unbounded, we could write
\begin{align*}
    \begin{pmatrix} \omega_i \\ \bm{b}_i \end{pmatrix} = 
    \bm{\mu}_R + \bm{L} \bm{z}_i
    =
    \begin{pmatrix}
        \mu_{\omega} \\ \mu_{\bm{b}}
    \end{pmatrix}
    +
    \begin{pmatrix}
        l_{11}      & \bm{0} \\
        \bm{l}_{21} & \bm{L}_{22}
    \end{pmatrix}
    \begin{pmatrix}
        z_{1i} \\ \bm{z}_{2i}
    \end{pmatrix}
    =
    \begin{pmatrix}
        \mu_{\omega} \\ \bm{\mu}_{\bm{b}}
    \end{pmatrix}
    + \begin{pmatrix}
        l_{11} z_{1i} \\ \bm{l}_{21} z_{1i} + \bm{L}_{22} \bm{z}_{2i}
    \end{pmatrix},
\end{align*}
where $\bm{z}_i = (z_{1i}, \bm{z}_{2i}') \sim N_4(\bm{0}, \bm{I})$. If we sample $\omega_i$ from its truncated normal distribution, we can obtain $z_{1i} = (\omega_i - \mu_{\omega}) / l_{11}$. It follows that, given $\omega_i$, we can generate $\bm{b}_i | \omega_i \sim N_3(\mu_{\bm{b} | \omega_i}, \bm{\Sigma}_{\bm{b} | \omega})$ by generating $\bm{z}_{2i} \sim N_3(0, \bm{I}_3)$ and applying the transformation
\begin{align}
    \bm{b}_i = \bm{\mu}_{\bm{b}} + \frac{\omega_i - \mu_{\omega}}{l_{11}} \bm{l}_{21} + \bm{L}_{22} \bm{z}_{2i}.
    \label{eq:btran}
\end{align}
This transformation in \eqref{eq:btran}, referred to as a non-centered parameterization, ``shifts'' the constraints from the (many) random effects to the (relatively fewer) parameters, which can improve the efficiency of the HMC sampler.

\subsection{Prior elicitation}
In this section, we discuss prior elicitation for our proposed model. We propose weakly informative priors for the random effects parameters and noninformative priors otherwise.

\subsubsection{The time-to-event distribution}
We assume a parametric proportional hazards (PH) model for the time-to-event outcome for the remainder of this paper. Write $\bm{\theta}_T = (\bm{\gamma}', \bm{\lambda}')'$, where $\bm{\gamma}$ is a vector of regression coefficients and $\bm{\lambda}$ is a vector of nuisance parameters pertaining to the baseline hazard. Let $\bm{w}_i = (w_{i1}, \ldots, w_{iq})'$ denote a vector of regression coefficients for the time-to-event model and write $\bm{\gamma} = (\gamma_1, \ldots, \gamma_q)'$. Then we may write the density for the time-to-event outcome as
\begin{align}
    f_T(t | \bm{\theta}_T) = h_{T0}(t | \bm{\lambda}) e^{\bm{w}_i'\bm{\gamma}}
    \exp\left\{ -H_{T0}(t | \bm{\lambda}) e^{\bm{w}_i'\bm{\gamma}} \right\},
\end{align}
where $h_{T0}(\cdot | \bm{\lambda})$ is a baseline hazard function and $H_{T0}(t | \bm{\lambda}) = \int_{0}^{t} h_{T0}(s | \bm{\lambda}) ds$ is the corresponding cumulative hazard function. A sensible prior for $\bm{\gamma}$ is $\bm{\gamma} \sim N(\bm{0}, \sigma^2_{\gamma} \times I_q)$, where $\sigma_{\gamma}$ is taken to be noninformative (e.g., $\sigma_{\gamma} = 10$). The priors for $\bm{\lambda}$ will in general depend on the model. 

In our simulations and data analysis, we use a Weibull model and, for the shape and scale parameters, we elicit independent non-informative half-normal priors with location parameter $\mu = 0$ and scale parameter $\sigma = 10$. In general, any parametric time-to-event model may be used, such as the flexible PH model with piecewise constant baseline hazards.

\subsubsection{The random effects distribution}
For the random effects, write $\bm{\theta}_R = (\bm{\mu}', \bm{\sigma}_R', \bm{\Gamma}_R')'$, where $\bm{\mu} = (\mu_{\omega}, \bm{\mu}_{\bm{b}}')'$ contains the mean of the random effects, $\bm{\sigma}_R = (\sigma_{\omega}, \bm{\sigma}_{\bm{b}})$ contains the standard deviations of the random effects, and $\bm{\Gamma}_R$ is a correlation matrix. Hence, we may write the distribution of the random effects as
$$
   f_{R|T}(\bm{r}_i | \bm{\theta}_R) = 
   \frac{\phi(\bm{r}_i | \bm{\mu}_R, \text{diag}(\bm{\sigma}_R) \bm{\Gamma}_R  \text{diag}(\bm{\sigma}_R))}{ \Phi\left( \frac{t_i^* - \mu_{\omega}}{\sigma_{\omega}} \right) - \Phi\left( -\frac{\mu_{\omega}}{\sigma_{\omega}} \right) },
$$
where $\text{diag}(\cdot)$ is a function that converts an $m\times1$ vector into an $m \times m$ diagonal matrix. 

Without a priori knowledge of the correlation between the random effects, we may elicit $\pi(\bm{\Gamma}_R) \propto 1$, i.e., the prior on the correlation matrix of the random effects is uniform over the space of positive definite correlation matrices. More generally, we can consider the density proposed by \cite{lewandowski_generating_2009}, which is given by $\pi(\bm{\Gamma}_R | \eta) \propto \lvert \bm{\Gamma}_R \rvert^{\eta - 1}$, which is uniform over the space of positive definite correlation matrices when $\eta = 1$. When $0 < \eta < 1$, the density has a trough at the identity matrix, whereas the identity matrix is the mode when $\eta > 1$. For the standard deviations, we elicit half-normal priors. The elicitation of the scale parameter can be problem-specific. For example, in our data application, it is unlikely that the marginal variance of the longitudinal outcome exceeds $1$. We thus elicit independent $N^+(0, 1)$ priors for the standard deviations of the random effects, where $N^+(m, s^2)$ denotes the half-normal distribution with location $m$ and scale $s$. \cite{gelman_prior_2006} recommends priors in the half-t family, for which the half-normal is a special case.

The means of the random effects require careful prior elicitation. In particular, note that if $\mu_{b_1} = \mu_{b_2}$ (i.e., if the mean of the random slope prior to the change point is equal to the mean of the random slope after the change point), the model becomes nonidentifiable. Put differently, a change point can only exist when the pre-slope and post-slope are different. We propose use of the generalized normal distribution (GND). The density function of the GND is given by
\begin{align}
    f(x | \mu, \alpha, \beta) \propto \exp\left\{ - \left( \frac{|x - \mu|}{\alpha} \right)^\beta \right\},
    \label{eq:gnd_density}
\end{align}
for $\mu \in \mathbb{R}$, $\alpha > 0$, and $\beta > 0$, which yields the normal kernel with variance $\alpha^2/2$ when $\beta = 2$. As $\beta \to \infty$, the density converges pointwise to a $U(\mu - \alpha, \mu + \alpha)$ distribution.

For example, as discussed in Section \ref{sec:motivation}, we have prior knowledge in our application that $\mu_{b_1} < 0$ and $\mu_{b_2} \ge 0$, i.e., that the tumor burden initially decreases before flattening out or increasing. Similarly, although the truncated normal distribution allows for the mean change point to be negative, it is reasonable to a elicit a prior that exhibits high mass for $\mu_{\omega} > 0$, which also facilitates interpretation since $\mu_{\omega}$ may be thought of as the mean change point on an untruncated scale as $t_i^* \to \infty$ for small $\sigma_{\omega}$. The GND thus allows us to elicit a prior that imposes soft thresholding adhering to this constraint, and is a strong justification for Bayeisan analysis of the model.

\subsubsection{The outcome distribution}
For the parameters pertaining to the outcome, we may elicit noninformative priors. We propose normal priors on the regression coefficients $\bm{\beta}$ and half-normal priors with a large prior variance on the standard deviation $\sigma$.

\section{Simulation study}
\label{sec:sims}
In this section, we present results on rigorous simulations for the proposed method. We conduct simulations across different sample sizes and censoring rates. We show that our proposed approach yields better results than a naive approach that does not consider bound constraints.

\subsection{Simulation procedure}
We now describe the data generation process for the simulation study. Unless noted otherwise, the parameters to generate data for the simulation study were selected to
mimic
the real clinical trial. These parameters are presented in Table~\ref{tab:para_pm}.

For $n \in \{ 100, 500 \}$, we generate covariates $\{ \bm{x}_i, i = 1, \ldots, n \}$ via bootstrap sampling from the pseudo data set. Progression times $\{ t_i^*, i = 1, \ldots, n \}$ were generated from a Weibull proportional hazards (PH) model, where the parameters were selected based on the analysis of the pseudo data set. The density for the Weibull PH with shape parameter $\alpha$, scale parameter $\eta$, and regression coefficients $\bm{\gamma}$ is given by $f(t_i^*) = \eta \alpha t_i^{* \alpha-1} \exp\{\bm{x_i'} \gamma - \eta t_i^{* \alpha} e^{\bm{x_i'} \gamma}\}$. Censoring times $\{ c_i^*, i = 1, \ldots, n \}$ are generated from an exponential distribution with rate parameter chosen so that approximately $q \times 100 \%$ of patients were censored, where $q \in \{0.20, 0.50\}$. The observed event time is $t_i = \min\{ t_i^*, c_i^* \}$, $i = 1, \ldots, n$. Random effects $\{ (\omega_i, \bm{b}_i), i = 1, \ldots, n \}$ are generated using the PTMVN($\bm{\mu}$, $\bm{\Sigma})$ distribution in \eqref{eq:ptmvn_joint} with $\omega_i \in [0, t_i^*]$.

For $i = 1, \ldots, n$ and $j = 1, 2, \ldots$, we generate visit times via $s_{ij}^* = | 0.1 \times j - z_{ij} |$, where $z_{ij} \sim \text{Half-Normal}(0, 0.02^2)$ so that $E[s_{ij}^*] \approx 0.10 \times (j-1) + 0.08$ years and $\text{Var}(s_{ij}^*) \approx 0.012^2$ years. The observed visit times for subject $i$, $s_{ij}$, is taken to be all visit times prior to the observed event time, i.e., we calculate $n_i = \max\{ j : s_{ij}^* \le t_i \}$ and take $s_{ij} = s_{ij}^*$ for $j = 1, \ldots, n_{i}$. If subjects progressed or dropped out prior to the first observed visit time, i.e., if $t_i < s_{i1}$, we set $s_{i1} = 0.1 \times s_{i1}^*$. The longitudinal outcome is then generated via
$
    y_{ij} \sim N\left( \bm{x}_i'\bm{\beta} + b_{0i} + b_{1i}(s_{ij} - \omega_i) + b_{2i}(s_{ij} - \omega_i), \sigma^2_y \right)
$
for $i = 1, \ldots, n$ and $j = 1, \ldots, n_i$.

We repeat this generative process for $B = 5000$ times. We compute bias, mean squared error (MSE), and 95\% credible interval coverage (Cover). For a parameter $\theta$, we compute these quantities as 
$\text{Bias} = B^{-1} \sum_{b=1}^B \left(E[\theta | D_b] - \theta \right)$, 
$\text{MSE} = B^{-1} \sum_{b=1}^B \left( E[\theta | D_b] - \theta \right)^2$, and 
$\text{Cover} = B^{-1} \sum_{b=1}^B I( L(\theta | D_b) \le \theta \le U(\theta | D_b) )$, where $D_b$ refers to the $b^{th}$ generated data set and where $L(\theta | D_b)$ and $U(\theta | D_b)$ are, respectively, the 2.5\% and 97.5\% posterior quantiles of $\theta$ for data set $D_b$.

We compare the performance of our approach with an approach that does not incorporate the bound of the random intercept. Specifically, we augment the approach of \cite{brilleman_bayesian_2017} to include a covariate in the longitudinal data, referring the resulting model as ``Longitudinal only''. The Longitudinal only model is represented hierarchically as
$
    (\omega_i, \bm{b}_i) \sim N_4(\bm{\mu}, \bm{\Sigma}), \ 
    y_{ij} \sim N\left( \bm{x}_i'\bm{\beta} + b_{0i} + b_{1i}(s_{ij} - \omega_i)I(s_{ij} \le \omega_i) + b_{2i}(s_{ij} - \omega_i) I(s_{ij} > \omega_i), \sigma^2_y \right),
$
which is similar to our model except it ignores the time-to-event model and the bound constraints. For both methods, $M = 10000$ posterior samples were taken after a burn-in period of 2000.

\subsection{Priors for the simulation study}
\label{sec:sim_priors}
The priors for the parameters pertaining to the time-to-event and longitudinal (excluding random effects) outcomes were taken to be noninformative. For the time-to-event outcome, we used priors $\gamma_1 \sim N(0, 10^2)$, $\eta \sim N^+(0, 10^2)$, and $\alpha \sim N^+(0, 10^2)$, where $N^+(m, s^2)$ is the half-normal distribution with location $m$ and scale $s$ and where $\eta$ and $\alpha$ are the baseline scale and shape parameters for the Weibull model, respectively. For the longitudinal parameters, we elicited $\beta_1 \sim N(0, 10^2)$ and $\sigma_y \sim N^+(0, 10^2)$.

Following previous approaches (e.g., \cite{brilleman_bayesian_2017}), we elicit weakly informative priors for the random effects to encourage identifiability and convergence of the MCMC chain. For the random effects, we assume $\mu_{\omega} \sim \text{GND}(0.5, 0.5, 8)$, $\mu_{b_0} \sim \text{GND}(0, 1, 8)$, $\mu_{b_1} \sim \text{GND}(-0.5, 0.5, 8)$, and $\mu_{b_2} \sim \text{GND}(0.5, 0.5, 8)$, where the GND density is provided in \eqref{eq:gnd_density}. These priors are approximately uniform over $\{ (\mu_{\omega}, \mu_{b_0}, \mu_{b_1}, \mu_{b_2}) : \mu_{\omega} \in [0, 1], \mu_{b_0} \in [-1, 1], \mu_{b_1} \in [-1, 0], \mu_{b_2} \in [0, 1] \}$. This prior is motivated by our data application since it can be safely assumed that the change point occurs within one year and that the pre-slope and post-slope are of different signs. Independent $\text{Half-Normal}(0, 1)$ priors were assumed for the standard deviations of the random effects. Finally, we elicited a uniform prior over the space of positive definite correlation matrices.

\begin{table}[ht]
\centering
\resizebox{\columnwidth}{!}{
\begin{tabular}{c|c|c|c|c|c|c|c|c|c}
    $\gamma_1$ & $\eta$ & $\alpha$ & $\beta_1$ & $\sigma_y$ & $\mu_{\omega}$ & $\bm{\mu_b}$ & $\sigma_{\omega}$ & $\bm{\sigma_b}$ & $\bm{\Gamma}$\\ \hline
    0.18 & 3.76 & 1.88 & -0.01 & 0.08 & 0.90 & $\left[\begin{array}{c}

    -0.50 \\
    -0.20\\
     0.60
        \end{array}\right]$ & 0.15 & $\left[\begin{array}{c}
    0.20\\
    0.27 \\
    1.20
        \end{array}\right]$ & $\begin{pmatrix}
     1.000 & -0.415& -0.220& -0.280 \\
    -0.415 & 1.000 & 0.560 & 0.200 \\
    -0.220 & 0.560 & 1.000 & 0.185\\
    -0.280 & 0.200 & 0.185 & 1.000
\end{pmatrix}
$
\end{tabular}}
\caption{Parameters used for the simulation study.}
\label{tab:para_pm}
\end{table}


\subsection{Results of the simulation study}
Results from the simulation study are presented in Table~\ref{tab:sims_20pctcensor} when approximately $20\%$ of subjects have censored event times and Table~\ref{tab:sims_50pctcensor} when subjects have approximately $50\%$ censored event times. For the joint model, increasing the sample size results in less bias, lower MSE, and 95\% credible interval (CI) coverage closer to the nominal level when the proportion of censoring is fixed. If the time-to-event data is ignored, increasing the sample size results in worse performance for a fixed level of censoring. This is intuitive since the joint approach incorporates a bound on each individual's change point and appropriately propagates the uncertainty of the bound for censored individuals, whereas the longitudinal model does not take into account such a bound. As expected, increasing the proportion of censoring in general yields larger MSE and sometimes bias. Across all metrics, the proposed model performs at least as well as the longitudinal only model.

As expected, posterior results concerning parameters after the change point (i.e., $\mu_{b_2}$ and $\sigma_{b_2}$) have subpar performance. This is because our simulations assume that individuals progress fairly quickly after their change point, resulting in few observations observed beyond the change point. When censoring is approximately 20\%, the bias decreases for the post-slope parameter ($\mu_{b_2})$ when $n = 100$ to $n = 500$, but the MSE does not change, which suggests that there is a high degree of uncertainty surrounding this parameter (Table~\ref{tab:sims_20pctcensor}). When censoring is approximately 50\%, the bias and MSE improves for the joint approach. The 95\% credible intervals (CIs) for $\mu_{b_1}$ contain the true value more than 95\% of the time. Coupled with the large MSE, this indicates that the credible intervals for this parameter are very wide. Although the 95\% CI is slightly better under the longitudinal only model for $n = 100$ and $20\%$ censoring, bias and MSE are notably worse.


The simulation results show that the 95\% credible interval coverage is above 95 percent for many of the random effect parameters. This is likely due to the fact that we used weakly informative priors for these parameters. However, we stress that model identifiability becomes an issue when noninformative priors are used. Specifically, if $\mu_{b_1} \approx \mu_{b_2}$, then the means of the individual-specific change points are the same, calling into question whether a change point truly exists. Web Appendix C, we present simulation results for $n = 500$ and $n = 2000$ with approximately $20\%$ censoring using less informative priors. The coverage probability is below the 95\% level for both sample sizes, but there is substantial improvement in the coverage probability when $n = 2000$. Overall, the simulation results suggest that a large sample size is required for reliable asymptotic inference, necessitating a Bayesian approach with somewhat informative priors.

Additional simulations studied the impacts of frequency (i.e., number of longitudinal measurements) and dependent censoring (via a Gaussian copula). As expected, more frequent longitudinal measurements generally resulted in higher efficiency, while dependent censoring resulted in higher bias and mean squared error for the time-to-event outcome. However, we found that the estimation for the longitudinal parameters was quite robust to model misspecification. We also compare our analysis of the time-to-event parameters with a model that ignores the longitudinal data, finding that the performances are essentially identical. The results of these simulation exercises are presented in Section D of the Supplementary Appendix.

\begin{table}[ht]
    \centering
    \resizebox{\columnwidth}{!}{
    \begin{tabular}{
    |c|c|c|c|c|c|c|c|c|c|c|c|c|
    }
    \hline
    & \multicolumn{6}{c|}{$n=100$} & \multicolumn{6}{c|}{$n=500$} \\ \cline{2-13}
    & \multicolumn{3}{c|}{Joint model} & \multicolumn{3}{c|}{Longitudinal only} & \multicolumn{3}{c|}{Joint model} & \multicolumn{3}{c|}{Longitudinal only} \\ \cline{2-13}
    & {Bias} & {MSE} & {Cover} & {Bias} & {MSE} & {Cover} & {Bias} & {MSE} & {Cover} & {Bias} & {MSE} & {Cover} \\ \hline
    $\beta_1$         & 0.001 & 0.018 & 95.6  & 0.013 & 0.024 & 90.4       & 0.000 & 0.008 & 95.1  & 0.013 & 0.015 & 68.8   \\ \hline
    $\sigma_y$        & 0.001 & 0.005 & 94.2  & 0.003 & 0.006 & 87.6       & 0.000 & 0.002 & 94.9  & 0.002 & 0.003 & 78.9   \\ \hline
    $\mu_{\omega}$    & -0.099 & 0.125 & 99.1 & 0.003 & 0.597 & 86.1       & -0.008 & 0.047 & 98.8 & -0.148 & 0.237 & 68.9  \\ \hline
    $\mu_{b_0}$       & 0.026 & 0.064 & 97.6  & 0.240 & 0.248 & 9.3        & 0.002 & 0.034 & 97.5  & 0.246 & 0.247 & 0.0   \\ \hline
    $\mu_{b_1}$       & -0.002 & 0.078 & 97.8 & -0.108 & 0.804 & 52.7      & -0.002 & 0.049 & 95.4 & -0.036 & 0.851 & 2.3  \\ \hline
    $\mu_{b_2}$       & -0.080 & 0.145 & 99.7 & 0.575 & 1.016 & 89.2       & -0.041 & 0.145 & 99.7 & 0.652 & 0.905 & 75.0  \\ \hline
    $\sigma_{\omega}$ & 0.058 & 0.102 & 96.7  & 0.119 & 0.198 & 79.6       & 0.001 & 0.019 & 97.7  & 0.088 & 0.111 & 27.8   \\ \hline
    $\sigma_{b_0}$    & 0.014 & 0.032 & 96.3  & 0.055 & 0.085 & 82.3       & 0.000 & 0.008 & 95.7  & 0.064 & 0.070 & 10.2   \\ \hline
    $\sigma_{b_1}$    & -0.018 & 0.072 & 94.0 & 0.165 & 0.641 & 91.1       & -0.003 & 0.025 & 94.3 & 0.129 & 0.586 & 91.6   \\ \hline
    $\sigma_{b_2}$    & -0.309 & 0.462 & 85.2 & 0.675 & 1.181 & 90.2       & -0.099 & 0.246 & 92.5 & 0.031 & 0.438 & 89.6  \\ \hline
    \end{tabular} 
    }
    \caption{Simulation results when approximately 20\% of subjects are censored.}
    \label{tab:sims_20pctcensor}
\end{table}

\begin{table}[ht]
    \centering
    \resizebox{\columnwidth}{!}{
    \begin{tabular}{
    |c|c|c|c|c|c|c|c|c|c|c|c|c|
    }
    \hline
    & \multicolumn{6}{c|}{$n=100$} & \multicolumn{6}{c|}{$n=500$} \\ \cline{2-13}
    & \multicolumn{3}{c|}{Joint model} & \multicolumn{3}{c|}{Longitudinal only} & \multicolumn{3}{c|}{Joint model} & \multicolumn{3}{c|}{Longitudinal only} \\ \cline{2-13}
    & {Bias} & {MSE} & {Cover} & {Bias} & {MSE} & {Cover} & {Bias} & {MSE} & {Cover} & {Bias} & {MSE} & {Cover} \\ \hline
    $\beta_1$         & 0.003 & 0.019 & 95.0  & 0.013 & 0.024 & 90.7       & 0.001 & 0.008 & 94.8   & 0.013 & 0.016 & 69.8   \\ \hline
    $\sigma_y$        & 0.001 & 0.006 & 95.0  & 0.005 & 0.008 & 86.1       & 0.000 & 0.003 & 94.6   & 0.002 & 0.003 & 84.3   \\ \hline
    $\mu_{\omega}$    & -0.177 & 0.201 & 99.0 & 0.763 & 1.666 & 83.7       & -0.038 & 0.067 & 99.0  & -0.178 & 0.328 & 73.9  \\ \hline
    $\mu_{b_0}$       & 0.053 & 0.087 & 95.3  & 0.204 & 0.258 & 39.4       & 0.010 & 0.039 & 98.8   & 0.250 & 0.251 & 0.1   \\ \hline
    $\mu_{b_1}$       & -0.003 & 0.081 & 98.6 & -0.043 & 0.596 & 64.1      & -0.002 & 0.060 & 97.1  & -0.085 & 0.752 & 16.1 \\ \hline
    $\mu_{b_2}$       & -0.122 & 0.166 & 99.3 & 0.316 & 0.874 & 92.1       & -0.053 & 0.123 & 100.0 & 0.673 & 1.055 & 83.3  \\ \hline
    $\sigma_{\omega}$ & 0.124 & 0.184 & 94.6  & 0.247 & 0.377 & 89.8       & 0.007 & 0.026 & 98.5   & 0.068 & 0.108 & 56.5   \\ \hline
    $\sigma_{b_0}$    & 0.026 & 0.044 & 97.2  & 0.032 & 0.072 & 94.6       & 0.001 & 0.010 & 96.7   & 0.058 & 0.076 & 40.3   \\ \hline
    $\sigma_{b_1}$    & -0.025 & 0.090 & 96.0 & 0.108 & 0.569 & 82.1       & -0.007 & 0.038 & 94.2  & 0.165 & 0.593 & 93.5   \\ \hline
    $\sigma_{b_2}$    & -0.456 & 0.565 & 80.6 & 1.272 & 1.671 & 92.0       & -0.188 & 0.345 & 91.5  & 0.230 & 0.808 & 87.2  \\ \hline
    \end{tabular} 
    }
    \caption{Simulation results when approximately 50\% of subjects are censored}
    \label{tab:sims_50pctcensor}
\end{table}

To conclude, our joint modeling approach is a notable improvement over an approach that does not take into account bound constraints on the change point. Across almost all parameters, bias, MSE, and CI coverage are improved compared to the longitudinal only model.

\section{Application to the EMPOWER study}
\label{sec:analysis}
\begin{figure}
    \centering
    \includegraphics[width=0.9\textwidth]{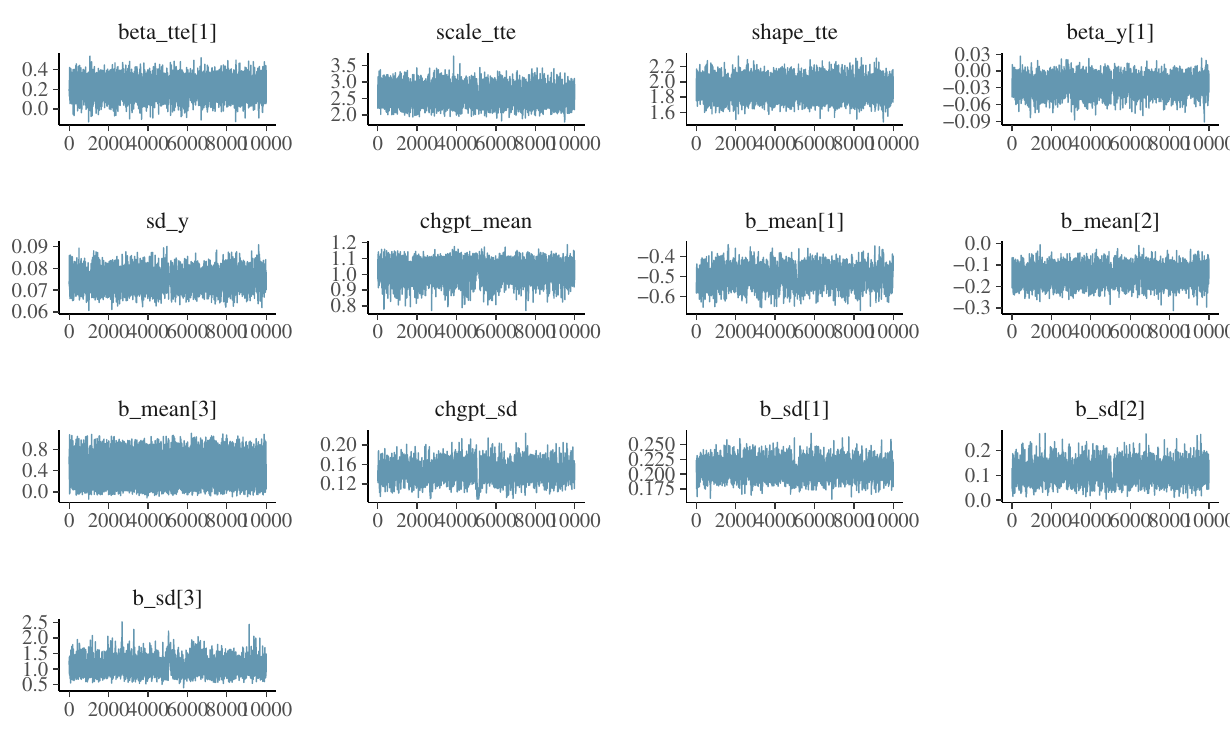}
    \caption{Trace plots of the model parameters for the simulated EMPOWER data.
    \texttt{beta\_tte[1]} = $\gamma_1$; 
    \texttt{scale\_tte} = $\eta$;
    \texttt{shape\_tte} = $\alpha$;
    \texttt{beta\_y[1]} = $\beta_1$;
    \texttt{sd\_y} = $\sigma_y$;
    \texttt{chgpt\_mean} = $\mu_{\omega}$;
    \texttt{chgpt\_sd} = $\sigma_{\omega}$
    \texttt{b\_mean[j]} = $\mu_{b_{j-1}}$ for $j = 1, 2, 3$;
    \texttt{b\_sd[j]} = $\sigma_{b_{j-1}}$ for $j = 1, 2, 3$;}
    \label{fig:traceplots}
\end{figure}

In this section, we provide analysis results on a simulated data set that mimics the control arm of the EMPOWER study. For brevity, we focus on inference for the longitudinal data model and random effects. The priors used were the same as in Section \ref{sec:sim_priors}. We generated 10,000 posterior samples after a burn-in period of 2000. Traceplots are presented in Figure~\ref{fig:traceplots}. The traceplots indicate that our MCMC sampler has converged to the stationary distribution.

Table~\ref{tab:posterior_ctrl} reports summary statistics from the posterior density. Note that if $\mu_{\omega} \pm 3\times \sigma_{\omega} \subset [0, \tau]$ for some $\tau > 0$, then $\mu_{\omega}$ is approximately equal to the mean change point for individuals with $t_i^* > \tau$, i.e., the mean change point under event times longer than $\tau$. Since the posterior mean of the change point location parameter is $E[\mu_{\omega} | D] = 1.011$ and that for the scale parameter is $E[\sigma_{\omega} | D] = 0.143$, we can interpret $\mu_{\omega}$ as the mean change point for individuals whose disease progression time is beyond $1.011 + 3 \times 0.143 = 1.440$ years. If the mean change point (say, $m_{\omega} = E[\omega_i]$) for the study population is of interest, it may be computed as
$
   m_{\omega} = \int E[\omega_i | t_i^* = t] f_T(t) dt
   = \mu_{\omega} - \sigma_{\omega} \int \left[  \frac{ \phi\left( \frac{t - \mu_{\omega}}{\sigma_{\omega}} \right) - \phi\left( \frac{t - \mu_{\omega}}{\sigma_{\omega}} \right) }{ \Phi\left( \frac{t - \mu_{\omega}}{\sigma_{\omega}} \right) - \Phi\left( \frac{t - \mu_{\omega}}{\sigma_{\omega}} \right) } \right] f_T(t) dt 
   ,
$
where we use Proposition~\ref{prop:ptmvn_mgf} and Equation \ref{eq:ptmvn_mean}. This integral may be computed using numerical or Monte Carlo integration plugging in the posterior samples for parameters.

\begin{table}[ht]
    \centering
    \begin{tabular}{
    |c|c|c|c|
    }
    \hline
    & {Mean} & {Standard Deviation} & {95\% Credible Interval}\\ \hline
    $\gamma_1$ &  $0.214$ & $0.091$ &  $(0.031, 0.388$)    \\ \hline
    $\eta$ &  $2.567$  & $0.246$ &   $(2.111, 3.082$)    \\ \hline
    $\alpha$ & $1.900$  & $0.118$   & $(1.675, 2.137$)     \\ \hline
    $\beta_1$ &   $-0.026$ & $0.015$ & $(-0.056, 0.003$) \\ \hline
    $\sigma_y$ & $0.075$ & $0.004$ & $(0.067, 0.083$)   \\ \hline
    $\mu_{\omega}$ & $1.011$  & $0.061$ &  $(0.874, 1.113$)     \\ \hline
    $\mu_{b_0}$ & $-0.498$ & $0.047$ & $(-0.589, -0.405$)     \\ \hline
    $\mu_{b_1}$ &  $-0.146 $& $0.038$ & $(-0.219, -0.071$)  \\ \hline
    $\mu_{b_2}$ &  $0.401$ & $0.243$  & $(0.016, 0.907$)  \\ \hline
    $\sigma_{\omega}$ & $0.143$ & $0.018$ &  $(0.104, 0.180$)   \\ \hline
    $\sigma_{b_0}$ & $0.207$ & $0.014$ &  $(0.182, 0.237$)   \\ \hline
    $\sigma_{b_1}$ & $0.113 $& $0.037$  & $(0.045, 0.188$)  \\ \hline
    $\sigma_{b_2}$ &  $1.076$ &  $0.248$ &   $(0.701, 1.689$)    \\ \hline
    \end{tabular}
    \caption{Posterior summary for the pseudo data set.}
    \label{tab:posterior_ctrl}
\end{table}

The pre-slope $\mu_{b_1}$ has posterior expectation $E[\mu_{b_1} | D] = -0.146$ with a 95\% credible interval of $(-0.219, -0.071)$. Thus, upon receiving the control of the EMPOWER study, tumor burden is expected to decline by 14.6\% per year. The post-slope $\mu_{b_2}$ has posterior expectation $E[\mu_{b_2} | D] = 0.401$. This indicates that, after the change point, tumor burden is expected to increase by approximately 40\% per year. Such a sharp increase implies a patient's cancer progresses rapidly after the change point. Since patients drop out of observation upon progression, there is not much data for tumor burden after the change point, hence why the posterior standard deviation $\text{SD}(\mu_{b_2} | D)$ is large and the credible interval is wide compared to the corresponding pre-slope parameters.

\begin{figure}
    \centering
    \includegraphics[width=0.9\textwidth]{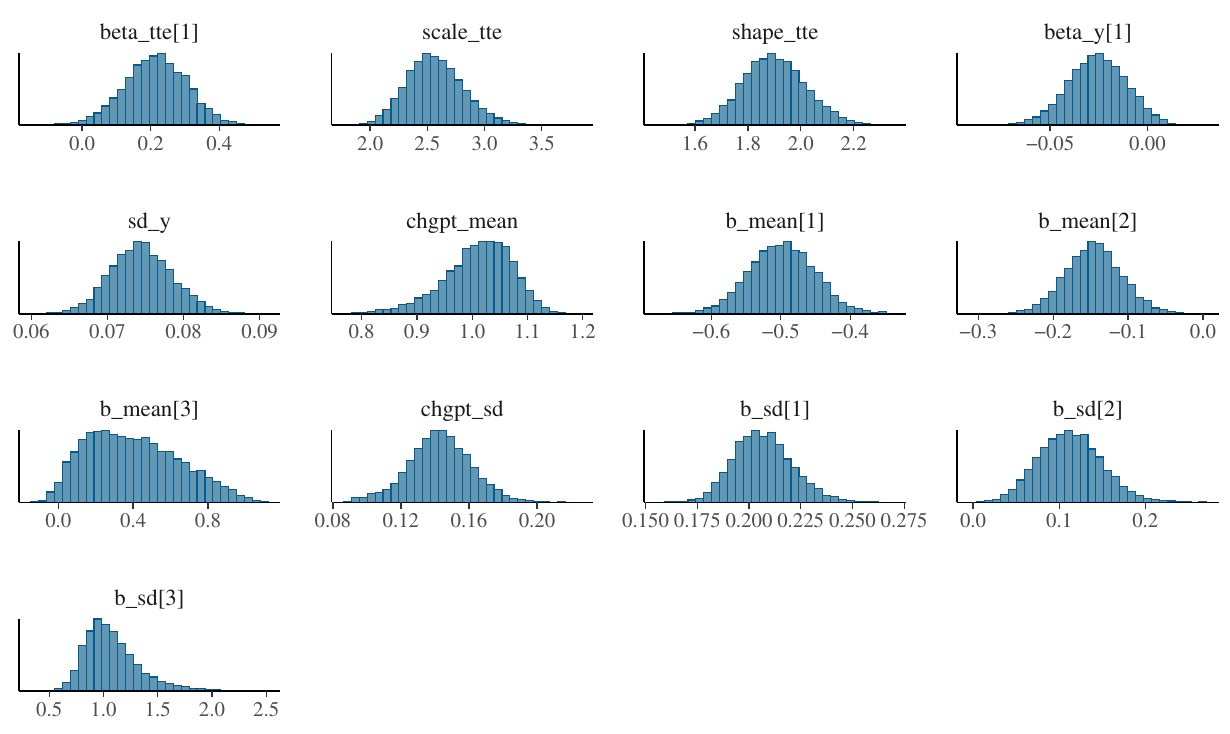}
    \caption{Histogram of the posterior samples for the simulated EMPOWER study data. 
    \texttt{beta\_tte[1]} = $\gamma_1$; 
    \texttt{scale\_tte} = $\eta$;
    \texttt{shape\_tte} = $\alpha$;
    \texttt{beta\_y[1]} = $\beta_1$;
    \texttt{sd\_y} = $\sigma_y$;
    \texttt{chgpt\_mean} = $\mu_{\omega}$;
    \texttt{chgpt\_sd} = $\sigma_{\omega}$
    \texttt{b\_mean[j]} = $\mu_{b_{j-1}}$ for $j = 1, 2, 3$;
    \texttt{b\_sd[j]} = $\sigma_{b_{j-1}}$ for $j = 1, 2, 3$;
    }
    \label{fig:posterior_hist}
\end{figure}

Figure~\ref{fig:posterior_hist} depicts histograms summarizing the posterior density of the parameters for the simulated EMPOWER study data. While some of the posterior parameters are approximately symmetric and bell shaped (e.g., $\gamma_1$, $\beta_1$), others are quite skewed (e.g., $\mu_{b_2}$, $\sigma_{b_2}$). Thus, a frequentist analysis relying on asymptotic distributions may not be appropriate for this model and data set. Note that the posterior of $\mu_{b_2}$ looks approximately uniform over $[0.0, 0.5]$. This is because there is not much data to estimate the slope of the longitudinal trajectory beyond the change point, and we imposed a prior that is approximately uniform over $[0, 1]$.

\begin{figure}
    \centering
    \includegraphics[width=0.9\textwidth]{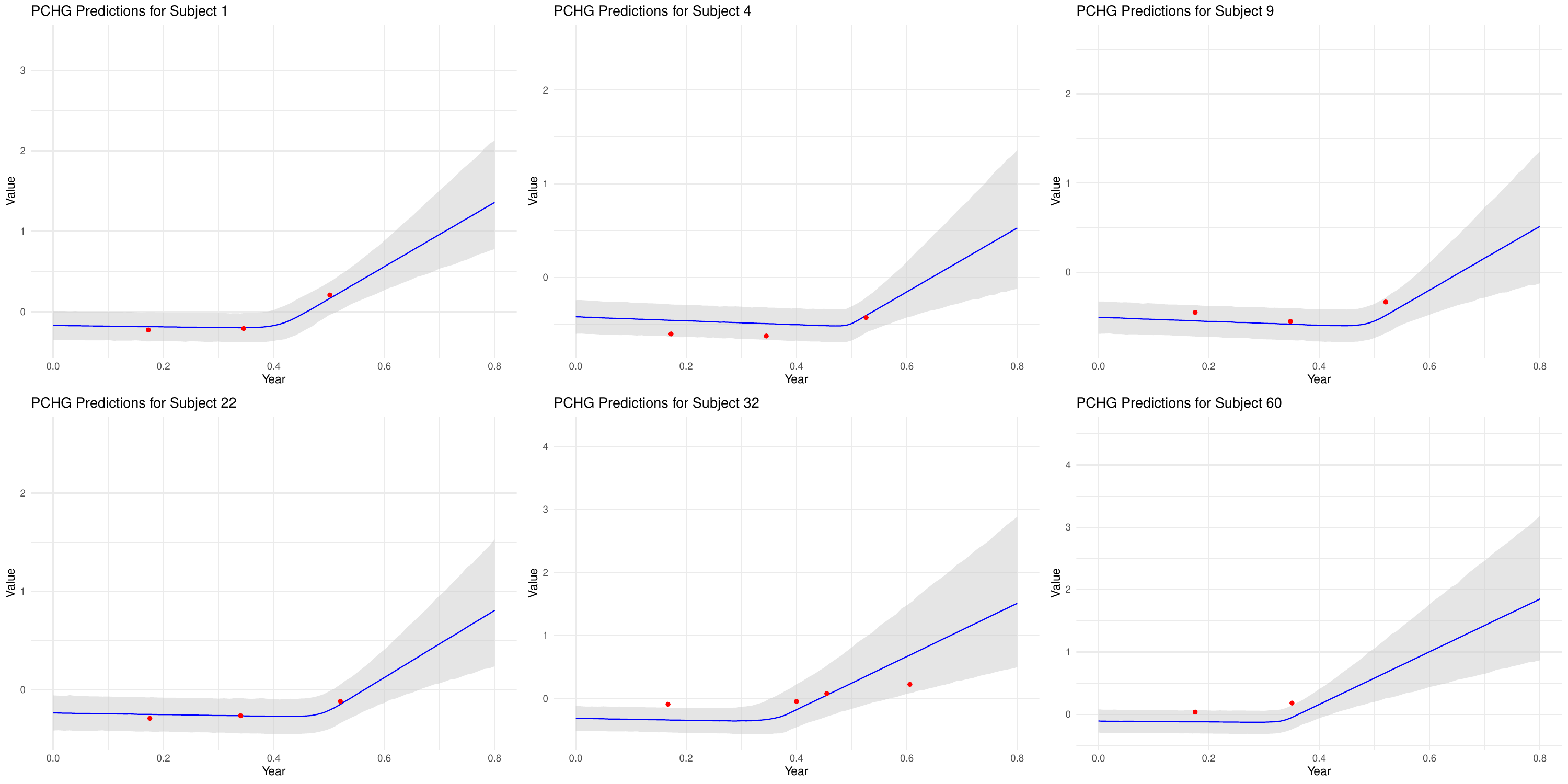}
    \caption{Posterior predictive plots for the longitudinal outcome for a subset of patients. The dots represent the observed values, and the line and shaded region depict the posterior predictive mean and 95\% prediction intervals, respectively.}
    \label{fig:posterior_check}
\end{figure}

When looking at the longitudinal outcomes in Figure~\ref{fig:spaghetti}, it is tempting to be concerned that our model may be misspecified since some individuals do not appear to be at risk of a change point (particularly those who experience the event very early). However, it is important to note that we do not observe these longitudinal measurements continuously, and the change point may have occurred between the baseline measurement and the first clinic visit. To provide evidence that this may be the case, we show posterior predictive plots in Figure~\ref{fig:posterior_check}. The plot shows that the observed longitudinal measurements fit largely within the 95\% prediction interval with few exceptions, including that for subject 60 who progresses rather quickly. Overall, the plots suggest that our model is not likely to be grossly misspecified. 

\section{Discussion}
\label{sec:discussion}
In this paper, we presented a novel model to jointly analyze time-to-event and longitudinal data when the random effects obey bounds depending on the event times. Simulation results indicated that the performance of our approach exceeds change point approaches that do not take into account the bounds. We applied our approach to the control arm of a real trial.

The methodology opens up several avenues for further research. First, an extension of the model to incorporate a cure fraction would be useful. Looking at the longitudinal outcomes among the censored subjects in Figure~\ref{fig:spaghetti}, a small fraction of the trajectories of tumor burden are flat, suggesting the possibility of cure. Moreover, thus far, the treatment arm of the  EMPOWER study has demonstrated such a significant benefit that many of these subjects have been administratively censored at the end of the study. When we applied our approach to the treatment arm, the posterior sampling had convergence issues. We conducted a time-to-event analysis with a cure fraction and found that over 30\% of observations were ``cured''. These individuals would never experience as change point and, as a result, our proposed model was misspecified when applied to the treatment arm.

Another potential avenue of research is the creation of clinical endpoints based on the random effects parameters. While it is tempting at first to consider the difference in the change point between the treated and control arms, consider that the magnitude and direction of the pre- and post-slope play a crucial role in determining the clinical benefit of the treatment. For example, suppose that the control arm has a mean change point occurring later than the treatment arm. Suppose further that the pre- and post-slopes for the treatment arm are both negative, while that for the control arm is initially negative but becomes positive. Looking at the difference in change points would lead one to conclude that the treatment is worse than control, which is clearly not the case here.

Finally, we mention that the proposed model can easily be extended to two arm setting in which appropriate estimand for that setting would then be defined that pertain to a treatment effect. The change point concept introduced here will allow us to consider important estimands for treatment effect evaluation, especially in the early stages of an oncology study. For example, we may use the difference in the mean change point time between treatment and control as the population-level summary of the endpoint (patient-level change point) to conclude if there is a treatment benefit in postponing the change point. Many other types of estimands can also be considered. The two arm extension with appropriate estimands is a topic of current research and development.

\bibliographystyle{unsrt}  
\bibliography{templateArxiv}

\end{document}